\documentstyle[12pt,aaspp4,epsf]{article}
\def\fun#1#2{\lower3.6pt\vbox{\baselineskip0pt\lineskip.9pt
  \ialign{$\mathsurround=0pt#1\hfil##\hfil$\crcr#2\crcr\sim\crcr}}}

\def\gap{\mathrel{\mathpalette\fun >}}

\lefthead{ }
\righthead{Resonant Orbits}

\begin{document}

\title{Resonant Orbits in Triaxial Galaxies}

\author{David Merritt and Monica Valluri}
\affil{Department of Physics and Astronomy, Rutgers University,
New Brunswick, NJ 08855}

\begin{abstract}
Box orbits in triaxial potentials are generically {\it thin},
that is, they lie close in phase space to a resonant orbit satisfying
a relation of the form $l\omega_1+m\omega_2+n\omega_3=0$ 
between the three fundamental frequencies.
Boxlets are special cases of resonant orbits in which one of the
integers $(l,m,n)$ is zero.
Resonant orbits are confined for all time to a membrane in 
configuration space; they play roughly the same role, in three 
dimensions, that periodic orbits play in two, generating families 
of regular orbits when stable and stochastic orbits when 
unstable.
Stable resonant orbits avoid the center of the potential; 
orbits that are thick enough to pass near the destabilizing
center are typically stochastic.
Resonances in triaxial potentials are most important at energies far 
outside the region of gravitational influence of a central black hole.
Near the black hole, the motion is essentially regular, although 
resonant orbits exist in this region as well, 
including at least one family whose elongation is 
parallel to the long axes of the triaxial figure.

\end{abstract}

\section {Introduction}

Motion in three-dimensional systems can differ in a number of 
qualitative ways from motion in one or two dimensions.
One example is complex instability (\cite{bro69}; \cite{pfe95}), 
in which 
perturbations of a periodic orbit diverge exponentially while 
rotating with a characteristic frequency.
Another example is Arnold diffusion (\cite{arn64}), the 
non-isolation of chaotic regions in phase space.

The focus of the present study is on a different property of
three-dimensional motion.
Associated with each degree of freedom of regular (quasi-periodic)
motion is a frequency $\omega_i$, 
the rate of change of the corresponding angle variable.
The character of the motion depends critically on whether the 
$\omega_i$'s are independent, or whether they satisfy one or more 
nontrivial linear relations of the form
\begin{equation}
\sum_{i=1}^N m_i\omega_i = 0
\label{res}
\end{equation}
with $N$ the number of degrees of freedom (DOF) and $m_i$ 
integers, not all of which are zero.
Generally there exists no relation like equation (\ref{res});
the frequencies are incommensurable,
and the trajectory fills its torus uniformly and densely 
in a time-averaged sense.
When one or more resonance relations are satisfied, however, 
the trajectory is restricted to a phase space region of 
lower dimensionality than $N$.

The importance of resonances for motion in two-dimensional 
systems is well known: resonant tori -- periodic, or closed, 
orbits -- are regions where the independent motions are coupled 
together, leading to a breakdown in perturbation expansions.
The character of the motion near a resonance is 
described by a number of theorems, including the 
Poincar\'e-Birkhoff and Floquet theorems.
When stable, resonant tori generate new families of regular orbits 
whose shape mimics that of the parent periodic orbit.
Unstable resonant tori are typically associated with 
a breakdown of integrability and with chaos.
In this sense, periodic orbits provide the phase space of a 2 DOF 
system with its structure.

In three dimensions, a single resonance relation like 
equation (\ref{res}) does not imply that an orbit will be closed; 
rather, it restricts the orbit to a space of dimension two.
An orbit satisying one such relation is therefore
``thin,'' confined for all time to a (possibly self-intersecting)
membrane.
In order for an orbit in a 3 DOF system to be
closed, it must satisfy two such independent relations; 
only then is the motion confined to a one-dimensional curve.
One expects that orbits satisfying two 
resonance relations will be rare compared to orbits satisfying 
just one, and hence that thin orbits -- rather than periodic 
orbits -- are the objects of fundamental importance in 
structuring the phase space of 3 DOF systems.
We present evidence in support of this hypothesis below.

The emphasis in the present paper is on three-dimensional 
resonances, i.e., resonances for which each of the 
integers $m_i$ in equation (1) is nonzero.
Thin orbits generated from a two-dimensional resonance have been 
widely studied; 
examples are the ``thin tubes'' discussed by Bishop, de Zeeuw 
and collaborators  
(\cite{bis87}; \cite{deh90}; \cite{ezl90}; \cite{zes96}).
Thin tube orbits exist even in fully integrable potentials; they 
are linked to the $1:1$ resonant orbits in the principal planes of 
St\"ackel models (\cite{dez85}).
By contrast, most of the resonant orbits discussed here can not 
be associated with any planar orbit; their existence is a consequence 
of a coupling 
between all three degrees of freedom and they have no analog in 
two-dimensional or in St\"ackel potentials.

The importance of resonant orbits in 3 DOF systems 
was suggested by a number of recent studies 
(\cite{caa98}; \cite{pal98}; \cite{vam98}; \cite{waf98})
that applied torus-construction machinery to motion in
triaxial potentials.
These studies noted that the phase space of box orbits, 
i.e. orbits with a stationary point, is densely structured by 
resonances, especially in models with central 
mass concentrations or nuclear ``black holes.'' 
Valluri \& Merritt (1998) noted that a resonance relation 
like equation (1) implies a reduction in the dimensionality of an orbit 
and presented an illustration of a thin box orbit.

Here we study the properties of such orbits in more detail.
Section 2 discusses the reduction in dimensionality that follows 
from the existence of a resonance between the three degrees of 
freedom.
Section 3 presents the properties of thin box orbits in a family 
of triaxial models with central density cusps and nuclear black 
holes; the emphasis is on the conditions under which such orbits 
are stable and can give rise to families of quasi-periodic 
orbits with finite thickness.
Section 4 extends this discussion to the central regions of a 
triaxial galaxy containing a black hole, where the potential is 
nearly Keplerian.
Some implications for galactic structure and evolution are 
presented in Section 5.

\section {Resonances}

In this section we discuss the effect of a resonance relation 
like equation (\ref{res}) on 
the motion of a regular orbit, and the differences between 
resonances in 2 and 3 DOF systems.

For a two-dimensional regular orbit with fundamental frequencies
$\omega_1$ and $\omega_2$, the angle variables are
\begin{equation}
\theta_1=\omega_1 t,\ \ \ \ \theta_2=\omega_2 t,
\label{reson}
\end{equation}
which define the surface of a torus.
The constants indicating phase on the torus have been set to 
zero without loss of generality.
Because of the quasi-periodicity of the orbit, its torus can be 
mapped onto a square in the $(\theta_1,\theta_2)$-plane, with 
each side ranging from $0$ to $2\pi$ (Figure 1);
the top and bottom of the square are identified with each other,
as are the left and right sides.
In the general case, the frequencies $\omega_1$ and $\omega_2$
are incommensurate and the trajectory densely covers the entire 
$(\theta_1,\theta_2)$-plane after an infinite time.
However if the ratio $\omega_1/\omega_2=|m_2/m_1|$
is a rational number, i.e. if $m_1$ and $m_2$ are integers,
the orbit closes on itself after 
$|m_2|$ revolutions in $\theta_1$ and $|m_1|$ revolutions in 
$\theta_2$ and fills only a one-dimensional subset of its torus
(e.g. \cite{arn73}, p. 164).
Its dimensionality in configuration space is also one.
Such an orbit has a single fundamental frequency 
$\omega_0 = \omega_1/m_2 = \omega_2/m_1 = 2\pi/T$, with $T$ the 
orbital period; after an elapsed time $T$, the trajectory returns 
to its starting point in phase space.

The effect of resonances on the motion in two-dimensional, 
non-integrable potentials is well understood 
(\cite{arn89}; \cite{lil92}).
Resonant orbits, when stable to perturbations, are 
associated with families of regular orbits, 
and when unstable generate regions of stochasticity.
Examples of resonant orbit families in two-dimensional 
galactic potentials are the boxlets 
(\cite{mis89}) which exist in the principal planes of triaxial 
models.
Two-dimensional resonances in the meridional plane of axisymmetric systems 
(\cite{eva93}) and in planar barred potentials 
(\cite{cog89}) have also been treated in detail.

In the case of a three-dimensional regular orbit, the angle
variables are
\begin{equation}
\theta_1=\omega_1 t,\ \ \ \ \theta_2=\omega_2 t,\ \ \ \ 
\theta_3=\omega_3 t.
\end{equation}
The orbit may now be mapped into a cube whose axes are 
identified with the $\theta_i$ (Figure 2).
If the $\omega_i$ are incommensurate, this cube will be 
densely filled after a long time.
However if a single condition of the form
\begin{equation}
m_1\omega_1 + m_2\omega_2 + m_3\omega_3 = 0
\label{commen}
\end{equation}
is satisfied, with the $m_i$ integers (not all of which are zero), 
the motion is restricted for all time to a two-dimensional subset of its torus 
(\cite{bor60}, p. 91; \cite{gol80}, p. 470).
An example is illustrated in Figure 2, with $(m_1,m_2,m_3) = 
(2,1,-2)$.
Such an orbit is not closed; instead, as suggested by 
Figure 2, it is {\it thin}, restricted to a 
sheet or membrane in configuration space, which it fills densely
after infinite time.

Just as in the two-dimensional case,
the condition (\ref{commen}) may be 
used to reduce the number of independent frequencies by one.
Defining the two ``base'' frequencies 
\footnote{The term ``base frequency'' is used here in a similar, but
more general, sense than in the study of Carpintero \& Aguilar (1998),
who considered only 2D resonances.}  
$\omega_0^{(1)}, \omega_0^{(2)}$ as
\begin{equation}
\omega_0^{(1)} = \omega_3/m_1,\ \ \ \ \omega_0^{(2)} = 
\omega_2/m_1,
\end{equation}
we may write
\begin{eqnarray}
\omega_1 & = & -m_3\omega_0^{(1)} - m_2\omega_0^{(2)}, \nonumber \\
\omega_2 & = & m_1\omega_0^{(2)}, \nonumber \\
\omega_3 & = & m_1\omega_0^{(1)}.
\end{eqnarray}
Since the motion is quasi-periodic, i.e.
\begin{equation}
{\bf x}(t)  = \sum_k {\bf X}_k\exp i
\left(l_k\omega_1 + m_k\omega_2 + n_k\omega_3\right)t,
\label{qp1}
\end{equation}
with $(l_k,m_k,n_k)$ integers, it will remain quasi-periodic when 
expressed in terms of the two base frequencies:
\begin{eqnarray}
{\bf x}(t)  & = & \sum_i {\bf X}_k\exp i
\left[\left(-l_k m_3+n_km_1\right)\omega_0^{(1)} + 
\left(-l_k m_2 + m_k m_1\right) \omega_0^{(2)}\right]t \nonumber \\
& = & \sum_k {\bf X}_k\exp i
\left({l_k}'\omega_0^{(1)} + {m_k}'\omega_0^{(2)}\right)t, \nonumber \\ 
& = & \sum_k {\bf X}_k\exp i
\left({l_k}'\theta^{(1)} + {m_k}'\theta^{(2)}\right), \nonumber \\ 
{l_k}' & = & -l_k m_3 + n_k m_1, \ \ \ \ \ {m_k}' = -l_k m_2 + m_k 
m_1, \nonumber \\
\theta^{(1)} & = & \omega_0^{(1)}t, \ \ \ \ \ \ \ \ \ \ \ \ \ \ \ \ \ \ 
\theta^{(2)} = \omega_0^{(2)}t.  
\label{qp2}
\end{eqnarray}
A Fourier transform of the motion will therefore consist of a set 
of spikes whose positions in frequency space can be expressed as 
linear combinations of just two frequencies.
The choice of base frequencies made here is clearly not unique, 
a consequence of the fact that the orbit is not closed.

A relation like equation (\ref{commen}) will be called a 
``resonance'' even though it does not imply that any frequency pair
can be expressed as a ratio of integers.
The integer vector $(m_1,m_2,m_3)$ is the order of the resonance;
the degeneracy of the resonance is defined as the number of independent 
resonance relations that are satisfied by the $\omega_i$.
In the case of twofold degeneracy, two independent resonance relations 
apply:
\begin{eqnarray}
& m_1\omega_1 + m_2\omega_2 + m_3\omega_3 & = 0, \nonumber \\
& n_1\omega_1 + n_2\omega_2 + n_3\omega_3 & = 0,
\label{degen}
\end{eqnarray}
and each frequency $\omega_i$ may be expressed as a rational fraction 
of any other:
\begin{equation}
{\omega_1\over\omega_3} = {m_2n_3 - m_3n_2\over m_1n_2 - m_2n_1} 
= {l_1\over l_3},\ \ \ \ \ 
{\omega_2\over\omega_3} = {m_3n_1 -m_1n_3\over m_1n_2 - 
m_2n_1} = {l_2\over l_3},
\end{equation}
with $(l_1,l_2,l_3)$ integers.
The motion is therefore periodic with a single base frequency 
$\omega_0=\omega_1/\l_1=\omega_2/l_2=\omega_3/l_3$
and the trajectory is closed.
In a system with $N$ degrees of freedom, $N-1$ such conditions 
are required for closure; only in the 2DOF case does a single 
resonance condition imply closure.
\footnote{J. D. Meiss (private communication) makes a distinction
between ``resonant,'' or closed, orbits and 
``commensurable'' orbits which satisfy fewer than $N-1$ 
relations like equation (\ref{commen}).}

The reduction of the dimensionality of an orbit in the presence 
of a resonance has been appreciated
at least since Einstein's 1917 paper on rules for classical
quantization (\cite{ein17}; \cite{per77}; \cite{cmm82}).
But the role of resonant orbits in structuring the phase space of
generic, three-dimensional systems is still not well understood.
Almost all discussions of resonances in the galactic dynamics
literature have focussed on restricted cases: 
on resonances between only two of the three degrees of freedom 
(e.g. \cite{caa98}; \cite{waf98}); 
on doubly-degenerate, i.e. closed, orbits 
(e.g. \cite{pfe84}; \cite{con86});
or on special potentials in which the motion is globally 
resonant, such as the Kepler potential.
The KAM theorem leads us to expect that -- 
in 3 DOF as in 2 DOF systems -- 
resonant tori are regions where 
perturbation expansions break down, leading to a change in
the local structure of phase space.
We expect thin orbits to play approximately the same role, 
in three dimensions, that periodic orbits play in two.

\section{Resonant Box Orbits in a Family of Triaxial Models}

We test this hypothesis by exploring the properties of 
thin orbits in one family of triaxial potentials.
The mass density law from which the potential was generated,
via Poisson's equation, has the form
\begin{equation}
\rho(m) = {(3-\gamma)M\over 4\pi
abc}m^{-\gamma}(1+m)^{-(4-\gamma)}, \ \ \ \ 0\le\gamma<3
\label{dehnen}
\end{equation}
with $M$ the total mass and $m=\left[(x/a)^2 + (y/b)^2 + 
(z/c)^2\right]^{1/2}$ the radius-like variable. 
This mass model is the generalization to triaxial geometry of the 
spherical family described by Dehnen (1993).
The potential and forces in the triaxial geometry
may be expressed in terms of one-dimensional
integrals (\cite{mef96}).
Dehnen's law has a power-law central density dependence which
approximates the observed luminosity profiles of early-type
galaxies and bulges
(\cite{cra93}; \cite{fer94}; \cite{mef95}; \cite{geb96}).

To this model was added a central point with mass $M_h$.
Here and below, units are adopted such that $a=G=M=1$; $M_h$ is 
therefore defined as the mass of the central object in units of 
the total galaxy mass.
Following the usual convention, the $x$- and $z$-axes are
identified with the long and short axes of the figure.

The character of the motion in triaxial Dehnen models, at radii 
well outside the gravitational radius of influence of the black 
hole (if present), has been discussed by Wachlin \& Ferraz-Mello (1998) 
and Valluri \& Merritt (1998).
As in those studies, orbits were integrated from various starting 
positions with zero initial velocity
for $\sim 10^2$ orbital periods 
and their motion analyzed using Laskar's (1988, 1990) 
algorithm, a Fourier technique for extracting the fundamental 
frequencies $\omega_i$ of a regular orbit with high precision.
The additional techniques described in Valluri \& Merritt (1998) 
were used to find the integers $(l_k,m_k,n_k)$ associated with 
each distinct peak $\omega_k$ in the frequency spectrum (equation 
\ref{qp1}).
For stochastic orbits, Laskar's technique gives an approximation, 
valid over the integration interval, of the true (continuous) 
spectrum.
Stochastic orbits were identified by integrating each trajectory 
for two contiguous time intervals and comparing the ``fundamental 
frequencies'' computed over each interval.
The change in the ``fundamental frequency'' associated with the 
largest amplitude term in the spectrum, $\Delta\omega$, was taken 
as a measure of the rate of stochastic diffusion in phase space
(\cite{las93}).

Figure 3 shows initial condition spaces for two triaxial Dehnen 
models, the first with $\gamma=0.5$ and $M_h=0$, the second with 
$\gamma=0.5$ and $M_h=0.003$.
Both models have $c/a=0.5$ and $b/a=0.791$.
The top frames (reproduced with slight modification from 
\cite{vam98}) show one octant of the equipotential surface, 
each located slightly within the half-mass radius of the model 
(or at ``shell 8'' in the notation of those authors).
On this surface, a grid of $\sim 10^4$ orbits were begun with 
zero velocity and integrated for 100 orbital periods.
The density of the gray scale is proportional to the logarithm of 
the stochastic diffusion rate as measured by $\Delta\omega$ over 
the integration interval.
Initial conditions for which the motion 
was found to be regular are shown in white.

The model of Figure 3a is close to integrable, with a 
finite central force and a moderate central force gradient.
Most of the significant stochasticity in this model is confined 
to initial conditions that lie between the short and intermediate
axes, the ``$Y$-$Z$ instability strip'' first 
described by Goodman \& Schwarzschild (1981).

Elsewhere on the equipotential surface, one sees a complex
network of intersecting resonance zones, some regular (white)
and some stochastic (dark).
The starting points of the thin orbits lie along the centers of 
these zones.
Several of the most important resonance zones in Figure 3a
are labelled by their defining integers.
Three of these -- the $(2,0,-1)$ ($x$-$z$ banana) resonance, 
the $(4,-3,0)$ ($x$-$y$ pretzel) resonance, 
and the $(3,0,-2)$ ($x$-$z$ fish) resonance -- 
are families that connect smoothly to periodic orbits in one of 
the principal planes, i.e. to ``boxlets'' 
(\cite{mis89}).
Others -- e.g. the $(3,-1,-1)$, $(2,1,-2)$ and $(4,-2,-1)$ resonances 
-- are not related to any planar periodic orbit; these resonances 
are characterized by nonzero values for each of the integers 
$m_i$.

Examples of thin orbits from each of these six families 
are shown in Figures 4 and 5.
Figure 4 plots intersections of the orbits with the three 
principal planes; because these orbits are thin, their 
intersection with any plane defines a curve or set of curves,
rather than a finite area as in the case of a volume-filling orbit.
None of the orbits passes precisely through the center although 
all of them come quite close.
Figure 5 presents views of the surfaces defined by the orbits.
These plots were generated 
using Laskar's algorithm to extract the frequency spectra,
equation (\ref{qp1}), followed by equation (\ref{qp2}) 
which yields the Cartesian coordinates 
in terms of the two reduced angle variables $(\theta^{(1)}, 
\theta^{(2)})$.
The resulting (numerical) functions ${\bf x}(\theta^{(1)}, 
\theta^{(2)})$ define a surface that was plotted via the 
{\it Mathematica} routine ``ParametricPlot3D.''

When projected against the principal planes, the thinness of 
these orbits is not readily apparent and it is likely that thin 
box orbits were seen but not identified as such in many earlier 
studies. 
A possible example is shown in Figure 6 of Levison \& Richstone (1987).

While there are initial conditions in Figure 3a that generate 
closed orbits -- orbits restricted to a single curve in 
configuration space -- the majority of regular orbits 
are identifiable only with a singly-degenerate
resonance zone.
Further evidence for this interpretation is provided by Figure 6, 
which shows the frequency spectra of two orbits, computed by
Fourier analysis of the $z$-component of the motion.
The first orbit is from the regular region that lies at the 
intersection of the $(2,1,-2)$ and $(4,-2,-1)$ resonance zones in 
Figure 3a.
The intersection of these zones defines a regular region of 
degeneracy two, associated with the closed, $5:6:8$ orbit at its center.
The second orbit is from the $(2,1,-2)$ resonance zone; 
this orbit is not obviously identified with any closed orbit.

Many of the lines in the spectrum of the first orbit, Figure 6a, 
lie precisely at integer multiples of a single base frequency, 
$\omega_k=n_k\omega_0^{(1)}$, with $\omega_0^{(1)} = 0.05997853$.
This frequency is close to the (single) frequency of the $5:6:8$ 
periodic orbit whose starting point lies nearby on the 
equipotential surface.
In addition, the spectrum of Figure 6a contains pairs of lines that are 
offset symmetrically from the primary lines, at frequencies of
$\omega_k\pm\omega_0^{(2)}$ and 
$\omega_k\pm\omega_0^{(3)}$, where
$\omega_0^{(2)} = 0.012007$ and 
$\omega_0^{(3)} = 0.016179$.
These two additional frequencies may be interpreted as resulting from 
the slow libration, in two independent directions, 
of the orbit around the parent closed orbit.
(Binney \& Spergel (1982) motivate this interpretation in 
the context of a two-dimensional orbit.)
The spectrum of the orbit in Figure 6a is thus clearly 
recognizable as that of a perturbed, closed orbit.

By contrast, the spectrum of the second orbit (Figure 6b)
contains lines at integer multiples of {\it two} base 
frequencies.
These may be defined as 
$\omega_0^{(1)} = \omega_x = 0.29655554$, 
the frequency associated with the 
strongest line in the $x$-spectrum, and 
$\omega_0^{(2)} = \omega_z = 0.48353497$, 
the primary frequency of the $z$-motion.
The strongest line in the $y$-spectrum lies at $\omega_y = 
-2\omega_x + 2\omega_y = -2\omega_0^{(1)} + 2\omega_0^{(2)}$,
consistent with the location of this orbit within the $(2,1,-2)$ 
resonance zone.
If the orbit whose spectrum is shown in Figure 6b were precisely thin, 
all of its lines would be representable in terms of the two base frequencies, 
as discussed above.
However one observes a third frequency in the form of multiplets 
at frequencies
$l_k\omega_0^{(1)} + m_k\omega_0^{(2)} \pm \omega_0^{(3)}$, 
with $\omega_0^{(3)}=0.010803$.
The ``splitting'' frequency $\omega_0^{(3)}$ is the same
(within numerical precision) for all of the multiplets of Figure 
6b and in all three $(x,y,z)$ of the spectra.
Thus, this orbit is clearly identifiable as a perturbed thin 
orbit rather than as a perturbed closed orbit.

Both spectra show additional, higher-order multiplets at low 
amplitude.

Examination of the spectra of a larger set of orbits suggests 
that regular orbits whose starting points lie within a 
(singly-degenerate) resonant zone always have spectra like that 
of Figure 6b, i.e. with two base frequencies and a single 
splitting frequency,
rather than like that of Figure 6a, with
a single base frequency and two splitting frequencies.
In this sense it is reasonable to state that the majority 
of regular orbits at this energy are associated with thin orbits 
and not with closed orbits.

A striking feature of Figure 3a is the large number of distinct, 
and fairly narrow, resonance zones.
The reason for the narrowness of the zones is suggested by Figure 3c, 
which shows the distance of closest approach to the potential center of 
a set of orbits whose initial conditions lie along the heavy 
curve in Figure 3a.
As one passes through a stable resonance zone, the orbital pericenter 
distance reaches a maximum on the resonance, where the orbit has 
zero thickness.
Initial conditions that lie to either side of the resonance 
produce orbits with a finite thickness; as this thickness 
increases, the pericenter distance falls, and eventually the 
orbit becomes thick enough to pass through the center of the potential.
Orbits with pericenter distances close to zero are generally 
stochastic, as shown in Figure 3e -- a likely consequence of the 
steepness of the force gradient near the center, which causes the 
trajectory to become sensitive to small perturbations.
The precise distance from the center at which stochasticity sets 
in is different for each resonant family but is typically of 
order $\sim 0.005$ at this energy in this potential.
(The linear scale of the orbits is of order $1$ at this energy).
The narrowness of the resonance zones is therefore a consequence 
of the fact that only a slight offset of an orbit's starting 
point from resonance is sufficient to force it into the 
destabilizing center.

The space between the primary resonance zones in Figure 3a 
appears to be criss-crossed by a large number of narrower zones.
Some of these additional zones are identified in Figure 3c with their 
integer vectors $(m_1,m_2,m_3)$.
A further illustration of the dense packing of resonance zones is 
given in Figure 7, which shows the transition between the 
$(2,1,-2)$ and $(4,-2,-1)$ resonant orbits along the heavy line of 
initial conditions shown in Figure 3a.
As one moves between these two primary zones, one passes over the
two subsidiary zones -- with orders $(-5,4,0)$ and $(1,3,-3)$ -- 
which can also be clearly seen in the pericenter plot, Figure 3c.
But resonances of even higher order are apparent in Figure 7.
The only orbits in that figure that appear to be genuinely 
volume-filling -- numbers 11 and 16 -- are slightly stochastic, 
although weakly enough that their stochasticity does not allow them 
to visit the full volume defined by the equipotential surface 
over an integration interval of $\sim 10^2$ oscillations.
(The weak stochasticity is a consequence of the low degree of 
central concentration of this model, $\gamma=0.5$, and the 
absence of a central ``black hole.'')
Thus, essentially every regular orbit in this 
set appears to be associated with a resonance.

The denseness of the resonance zones is not surprising.
In an integrable potential, where the fundamental frequencies 
vary smoothly with initial conditions, it is well known that
resonant tori are dense in the phase space, 
just as rational numbers are dense in the space of real numbers.
Even a slight perturbation of the potential away from exact integrability 
would be expected to radically change the motion in the
neighborhood of each of these resonances, in the same way that
the motion in perturbed 2D systems is strongly affected by the 
existence and stability of closed orbits.

Some of the resonance zones in Figure 3a are present also in 
Figure 3b, which shows an equipotential surface in the model 
with an added central point mass, $M_h=0.003$.
However the higher-order resonance zones (e.g. $(6,-6,1)$, 
$(-5,4,0)$) have disappeared; the motion in the corresponding 
regions is now chaotic.
A second difference is that -- within a given resonance zone -- 
chaos sets in well before an orbit is thick enough to sample the 
center (Figure 3d, f).
Evidently, an added central mass point can induce stochasticity 
even in orbits that do not pass particularly close to the center.
Figure 8 shows the approximate pericenter distance at which 
integrability is destroyed for three of the resonant orbit 
families in Figure 3b, as a function of the central mass $M_h$; 
in each of these potentials, the amplitude of the long-axis orbit 
is about 2.
When $M_h$ exceeds $\sim 1\%$ the mass of the galaxy -- typical 
of the black holes in a number of early-type 
galaxies (\cite{for98}) -- a pericenter distance of $\sim 5\%$ of the 
orbital amplitude is sufficient to induce stochasticity for each 
of these families.
The highest-order resonant family in Figure 8, the $(4,-2,-1)$ 
family, has been rendered completely chaotic for $M_h\gap 
0.01$, and the other families disappear for $M_h\gap 0.03$.

Valluri \& Merritt (1998) reported a transition to global 
stochasticity in the phase space of box orbits when the mass of a 
central point exceeded $\sim 1-3\%$ the galaxy mass.
Figure 8 suggests a simple explanation for this transition:
when the central mass is sufficiently great, even resonant orbits 
are unable to avoid the center by a wide enough margin to remain 
stable.

It was argued above that the majority of regular orbits in these 
potentials are properly associated with thin orbits rather than 
with closed orbits.
Further evidence in support of this claim is presented in Figure 
9, which shows the variation of the splitting frequency 
$\omega_0^{(3)}$ defined above
with initial conditions as one moves across the 
$(2,1,-2)$ resonance zone in the potential with $\gamma=0.5$ and 
$M_h=0.0003$.
The variation in $\omega_0^{(3)}$ is generally smooth, 
peaking on the resonance and falling off to either side.
This smooth variation suggests a continuous dependence of orbital 
properties on initial coordinates near the resonance.
One also sees some discontinuities;
inspection of the individual orbits reveals the existence of 
additional resonances, i.e. closed orbits, at these points.
These closed orbits must in some sense be dense 
in phase space.
However the generally smooth variation of $\omega_0^{(3)}$ with 
initial conditions suggests that the majority of orbits in a 
singly-degenerate resonance zone can be usefully associated 
with the resonance.

\section {Orbits near the Central Black Hole}

The character of the orbits, as well as the relative importance 
of thin orbits, might be expected to change as one approaches the 
center of a triaxial potential containing a nuclear black hole.
Pfenniger \& de Zeeuw (1989) and Sridhar \& Touma (1998) 
noted that the motion in the neighborhood of the black hole, i.e. 
at radii such that the mass of the black hole is comparable to 
or greater than the enclosed mass in stars, should be 
approximately integrable, and these authors presented the 
results of numerical integrations in two-dimensional 
harmonic-oscillator potentials with added central point masses.

We begin by describing how the population of box orbits 
changes as one moves from the half-mass radius into the region 
where the forces are dominated by the black hole.
We chose a Dehnen model with $\gamma=0.5$ and $M_h=0.003$; the
latter value is typical of the black hole mass ratio in 
early-type galaxies (\cite{for98}), and the weak cusp 
specified by $\gamma=0.5$ is characteristic of bright elliptical 
galaxies (\cite{geb96}), in which the evolution 
timescales due to chaotic mixing are long enough that a triaxial 
shape might maintain itself for roughly a Hubble time 
(\cite{meq98}; \cite{vam98}).
Following Merritt \& Fridman (1996), the mass model without the 
black hole was divided into 21 ellipsoidal shells of equal mass; 
thus the outer edge of shell 1 contains 1/21 of the total mass 
and shell 21 lies at infinity.
The energy corresponding to each shell was defined as the value of 
the gravitational potential -- now including the contribution 
from the central black hole -- on the $x$-axis at the outer edge of 
the shell.
The ratio of $M_h$ to the stellar mass enclosed by shell I,
$M_h/M_I$, is then $21 \times 0.003/I = 0.063/I$.

At shell 8 ($M_h/M_I=0.0079$), the diffusion rate map is given by 
Figure 3b.
As the energy is reduced, the fraction of the starting points
on the equipotential surface that generate regular motion drops; 
by shell 2 ($M_h/M_I=0.031$), the motion is almost entirely 
chaotic, containing only small regular regions associated with the 
$(2,0,-1)$, $(3,0,-2)$ and $(3,-1,-1)$ resonances.
At shell 1 ($M_h/M_I=0.063)$ the motion of box orbits is 
essentially fully stochastic.

This transition to global stochasticity in the phase space
of box orbits as the energy is reduced is similar to the 
change observed by Valluri \& Merritt (1998), at a {\it fixed} 
energy (shell 8), as $M_h$ was increased.
Those authors found that the motion of box orbits
became almost completely stochastic when the mass of the black hole 
increased past $\sim 0.01$ times the total mass of the model, 
or equivalently, $\sim 0.02$ times the enclosed mass.
We speculate that a transition to global stochasticity in the 
phase space of box orbits generically occurs at radii where the 
black hole contains of order $\sim 1-3\%$ of the enclosed mass 
in triaxial potentials.

Inside of this radius, a zone of chaos was found to extend 
inwards, roughly to the radius at which the gravitational force 
from the black hole begins to dominate the force from the stars.
To explore this central region, we defined a new set of shells, 
denoted by the index $J$, which divide the previously-innermost 
shell ($I=1$) again into 21 equal-mass shells; thus shell $J=1$ 
encloses $(J/21)\times (1/21) = 0.0023$ of the total stellar mass, 
etc.
The ratio of the black hole mass to the enclosed stellar mass 
within this region is therefore $M_h/M_J \approx 1.323/J$.

Figure 10 illustrates the behavior of the box orbits as one 
moves from the inner edge of the zone of chaos, at shell 
$J=7$ ($M_h/M_J\approx 0.189)$, into shell $J=4$ ($M_h/M_J\approx 
0.331$).
In addition to plots of the diffusion rate, as in Figure 3, we 
show the ``frequency maps'' defined by Papaphilippou \& 
Laskar (1998).
The first regular orbits to appear inside of the zone 
of chaos are associated with the ($2,0,-1)$ resonance, the 
$x-z$ banana orbit. 
Unlike the banana orbit at higher energies, which lie close to 
the long ($x$) axis, this banana orbit has its stationary point near the short 
($z$) axis of the model. 
As one moves inward, a regular region appears around the short 
axis, and grows to include most of the equipotential surface, 
with the exception of a strip connecting the $x$ and $y$ axes.
A number of stable and unstable resonances can be seen at each 
shell but most of these are important only over a very narrow 
range of energies; the only stable resonance that persists over a 
signficant radial range is the ($1,-2,1$) resonance.
The rapid variation in the resonance zones as the energy is reduced 
corresponds to a shift in the fundamental frequencies toward
$\omega_x\approx \omega_y \approx \omega_z$,
the expected behavior as one approaches the Keplerian 
potential of the black hole.

We note that passage through the ``zone of chaos'' has the effect of 
reversing the dynamical roles of the long and short axes.
At large energies, motion is stable (unstable) in the vicinity of 
the long (short) axes, respectively, and an instability strip extends from the 
short to the intermediate axes (Goodman \& Schwarzschild 1981; 
Figure 3).
Inside the zone of chaos, motion near the {\it short} axis becomes 
stable, and the instability strip extends from the intermediate 
to the long axes (Figure 10).
As noted above, the $x-z$ banana orbit also changes its direction 
of elongation, from the long to the short axis.
Sridhar \& Touma (1998) noted, in their study of 
two-dimensional motion in the vicinity of a black hole,
that the regular orbits are often elongated in the direction of the 
shorter of the two axes; our results suggest that the same is
true with regard to the minor axis of a triaxial galaxy, at least
in the case of the particular family of potentials investigated 
here.

The two most important families of boxlike orbits near the black 
hole are illustrated in Figure 11.
The first family (Figure 11a) consists of regular orbits not associated 
with any low-order resonance; similar orbits in two dimensions were called 
``lenses'' by Sridhar \& Touma (1998).
Those authors noted that 2D lens orbits could be 
approximately described as precessing Keplerian ellipses  
with one focus on the black hole.
The orbits found here appear to be straightforward 
generalizations, to three dimensions, of Sridhar \& Touma's
lenses,
precessing independently in the $y-z$ and $x-z$ planes.
These orbits are essentially volume-filling although close 
inspection of their cross-sections reveals that many are 
associated with a high-order resonance.

The second major family consists of orbits 
associated with the ($1,-2,1$) thin box mentioned above.
The resonant orbits that give rise to this family 
(Figure 11b) are elongated 
parallel to the $x-y$ plane; the symmetric objects formed by 
superposition of four such orbits, reflected about the principal 
planes, might be useful for self-consistently reconstructing 
a triaxial figure.
The (symmetrized) lens orbits are elongated parallel to the short 
axis of the triaxial figure and would probably not be very useful 
for this purpose, as noted by Sridhar \& Touma (1998).

\section{Summary and Discussion}

Our principal conclusions follow.

1. Resonant orbits -- orbits confined for all time to a membrane 
in configuration space -- play roughly the same role, in three 
dimensions, that periodic orbits play in two, generating families 
of regular orbits when stable and stochastic orbits when 
unstable.

2. Box orbits in realistic triaxial potentials are generically 
thin.
Their thickness is limited by the requirement that 
they avoid the destabilizing center of the potential; orbits that
pass too near the center become stochastic.
In triaxial mass models with central ``black holes'' containing 
$\sim 0.1\% - 1\%$ of the total mass, box orbits become 
stochastic when their distance of closest approach to the center 
is $\sim 0.05-0.1$ times the half-mass radius of the model.

3. Resonances in triaxial potentials
are less important at energies where the gravitational 
potential is dominated by the central black hole.
In these central regions, most of the boxlike orbits are regular and 
associated with a single non-resonant family.
However resonant orbits still exist, including at least one 
family whose primary elongation is perpendicular to the 
short axis of the galaxy figure.

Resonances in triaxial potentials may be relevant to the following
problems of current interest.

1. The rates of many physical processes depend on the efficiency 
with which stars are supplied to the very center of a galaxy.
Examples are tidal disruption (\cite{ree92}) and accretion (\cite{nos83})
of stars by a black hole,
transfer of energy from stars to a binary
black hole (\cite{qui96}), interaction of stars with an accretion
disk (\cite{ost83}), etc.
In a triaxial galaxy, the rates of these processes would depend
sensitively on the fraction of stars associated with 
resonant orbits.

2. It is sometimes argued (e.g. \cite{dez96}) that
the effectiveness of central density cusps or black holes 
at inducing changes in the orbital structure of a triaxial 
galaxy should be lessened by figure rotation, because
the Coriolis force in a rotating potential tends to make the
box orbits ``centrophobic.''
As shown here, the regular box orbits even in a {\it non}rotating 
potential
are generically centrophobic; hence one might not expect figure
rotation to have any great effect on their behavior.
In fact a recent study (\cite{val99}) finds that 
stochasticity becomes {\it more} prevalent as the rate of figure
rotation is increased, 
apparently because the Coriolis force tends to broaden orbits
that would otherwise be thin, thus driving them into the center.

3. As illustrated in Figure 8, one can compute fairly precisely 
the distance of closest approach to the potential center at which 
a regular box orbit becomes stochastic.
This distance can be relatively large, of order $10\%$ the galaxy's 
half-mass radius, when the black hole contains $\sim 1\%$ the 
mass of the galaxy.
It follows that the effect of a central mass on 
the orbital structure of a triaxial galaxy need not be strongly 
dependent on the spatial scale of the mass distribution.
In fact, $N$-body simulations reveal that the response of an
initially triaxial galaxy to a central accumulation of mass is 
essentially independent of whether the mass is distributed
(e.g. \cite{kag91}; \cite{udr93}; \cite{dub94})
or concentrated in a point (\cite{meq98}).

\bigskip

This work was supported by NSF grant AST 96-17088 
and by NASA grant NAG 5-6037.
The authors are happy to acknowledge useful discussions with
A. Bahri, G. Contopoulos, J. Laskar, R. de la Llave, J. Meiss, 
J. Moser, R. Nityananda, and Y. Papaphilippou.

\clearpage

\figcaption[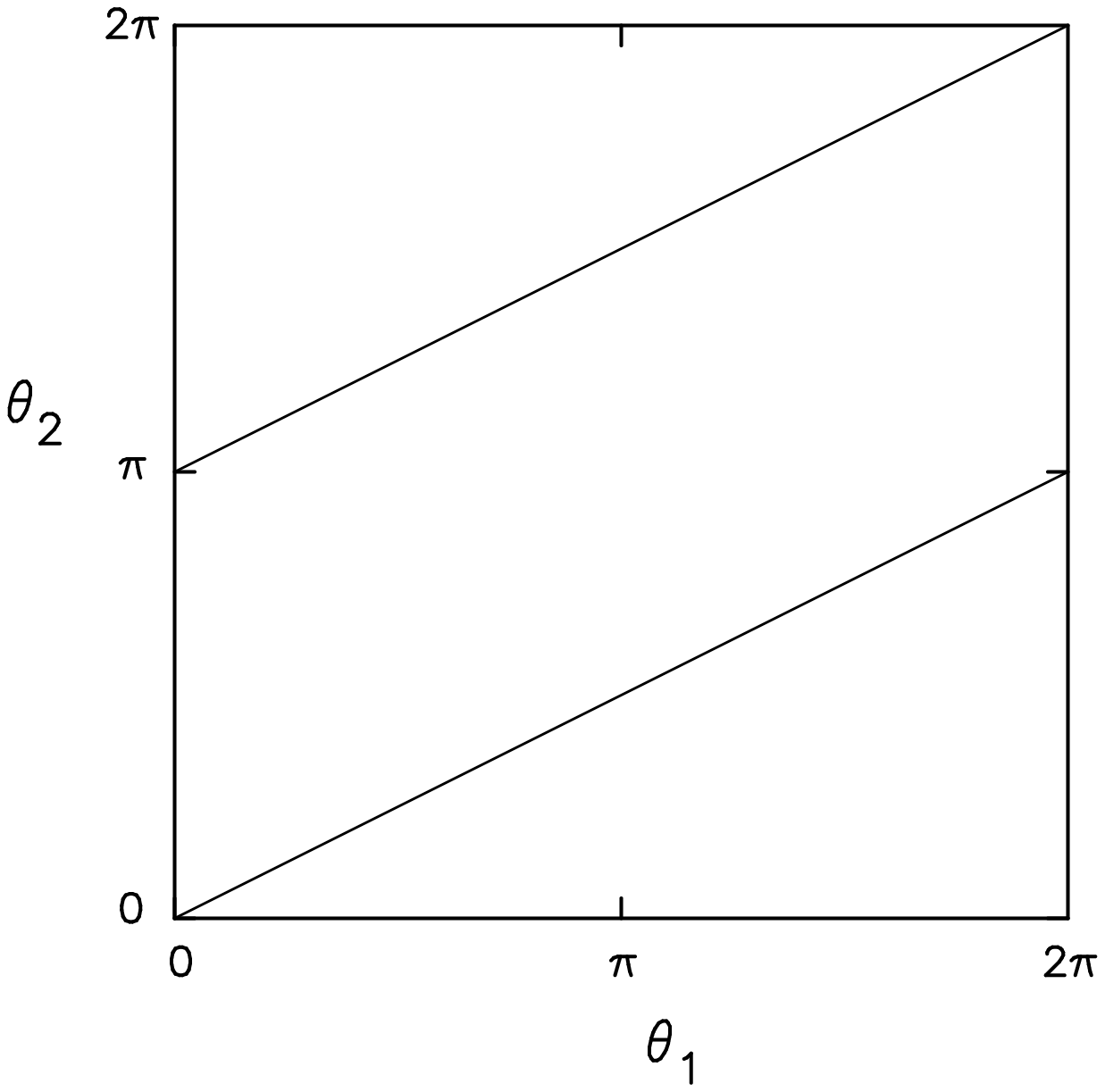]{\label{fig1}} 
A two-dimensional torus, shown here as a square with identified 
edges.
The plotted trajectory satisfies a $2:1$ resonance between the 
fundamental frequencies, $\omega_1 - 2\omega_2 = 0$.
The trajectory repeats after one rotation in $\theta_1$ and two 
rotations in $\theta_2$.
The corresponding orbit (e.g. a banana)
is closed in configuration space and confined to a 
one-dimensional curve.

\figcaption[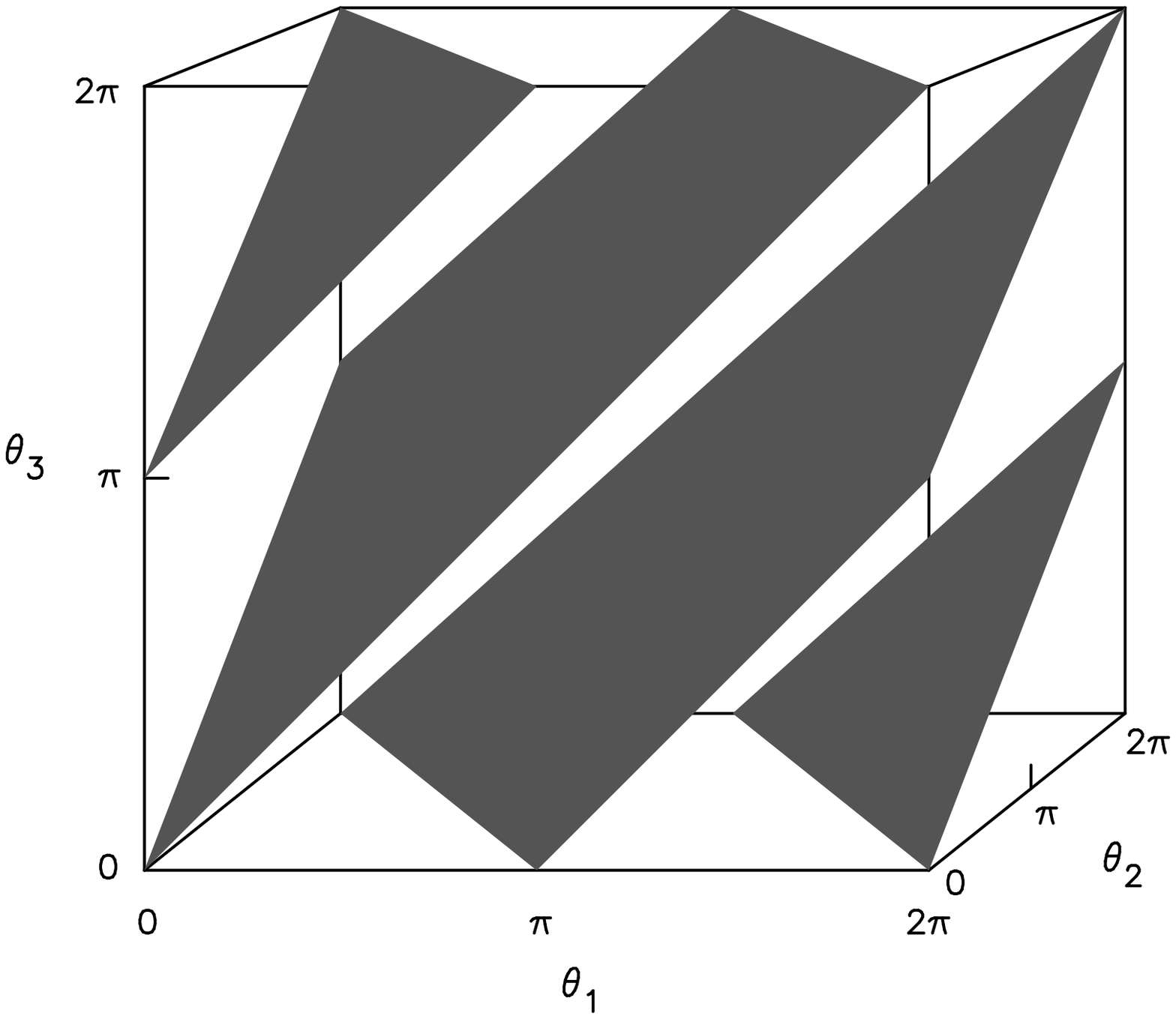]{\label{fig2}} 
A three-dimensional torus, shown here as a cube with identified 
sides.
The shaded region is covered densely by a resonant trajectory 
for which $2\omega_1 + \omega_2 - 2\omega_3 = 0$.
This trajectory is not closed, but it is restricted by the 
resonance condition to a two-dimensional subset of the torus.
The orbit in configuration space is thin, i.e. 
confined to a membrane.

\figcaption[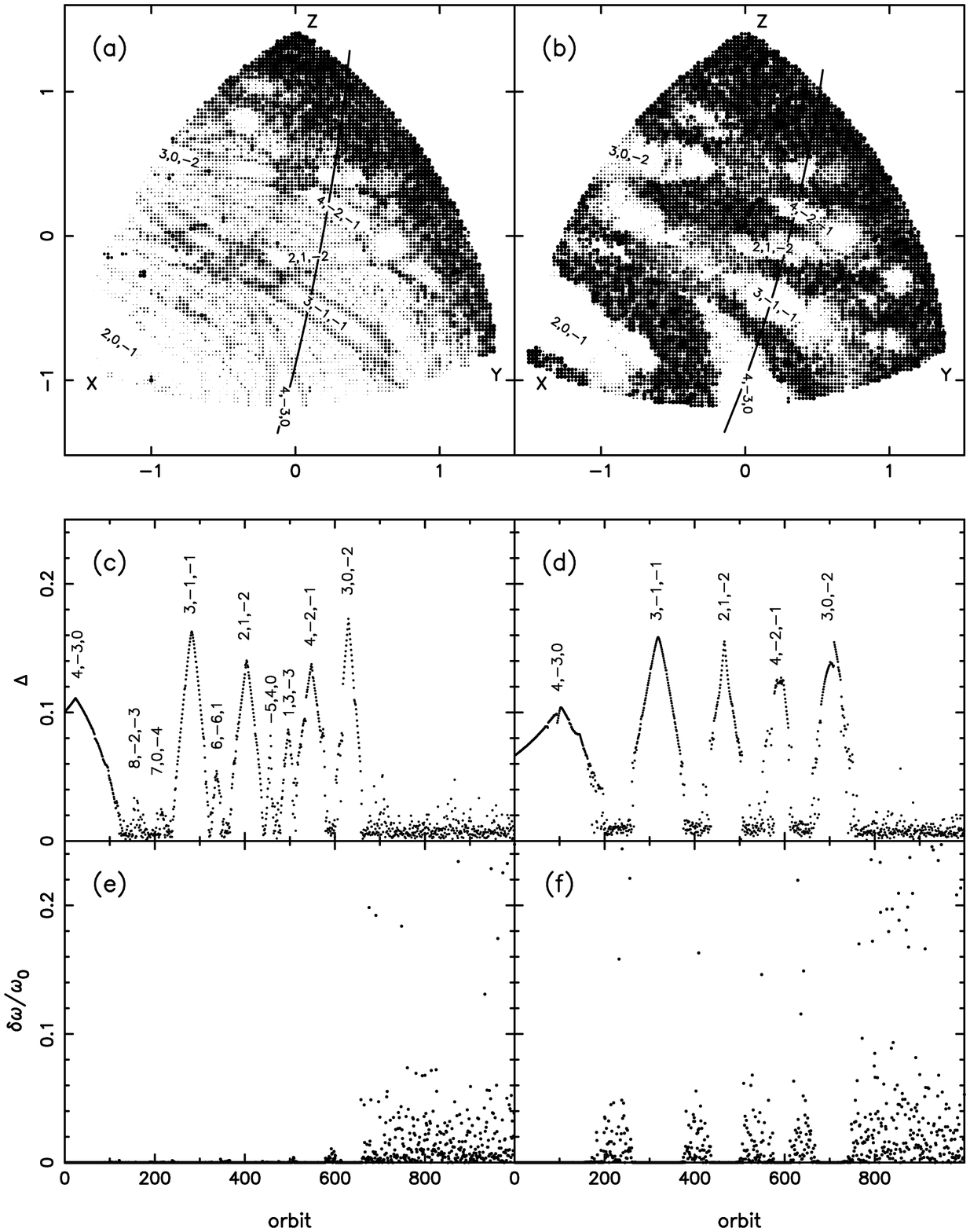]{\label{fig3}} 
Properties of box orbits in triaxial Dehnen models with 
$\gamma=0.5$.
Left panels: $M_h=0$; right panels: $M_h=0.003$.
(a), (b): One octant of the equipotential surface, on which 
orbits were started with zero velocity.
The top, left and right corners correspond to the $z$ (short),
$x$ (long) and $y$ (intermediate) axes.
The grey scale is proportional to the logarithm of the
diffusion rate of orbits in frequency space; initial conditions
corresponding to regular orbits are white.
The most important resonance zones are labelled with their 
defining integers $(m_1,m_2,m_3)$.
(c), (d): Pericenter distance $\Delta$ of orbits whose starting 
points lie along the heavy lines in (a) and (b).
The most important stable resonances are again labelled.
(e), (f): Degree of stochasticity of the orbits in (c) and (d), 
as measured by the change $\delta\omega$ in their ``fundamental 
frequencies.''
$\omega_0$ is the frequency of the long-axis orbit.
Regular orbits have $\delta\omega/\omega_0=0$.

\figcaption[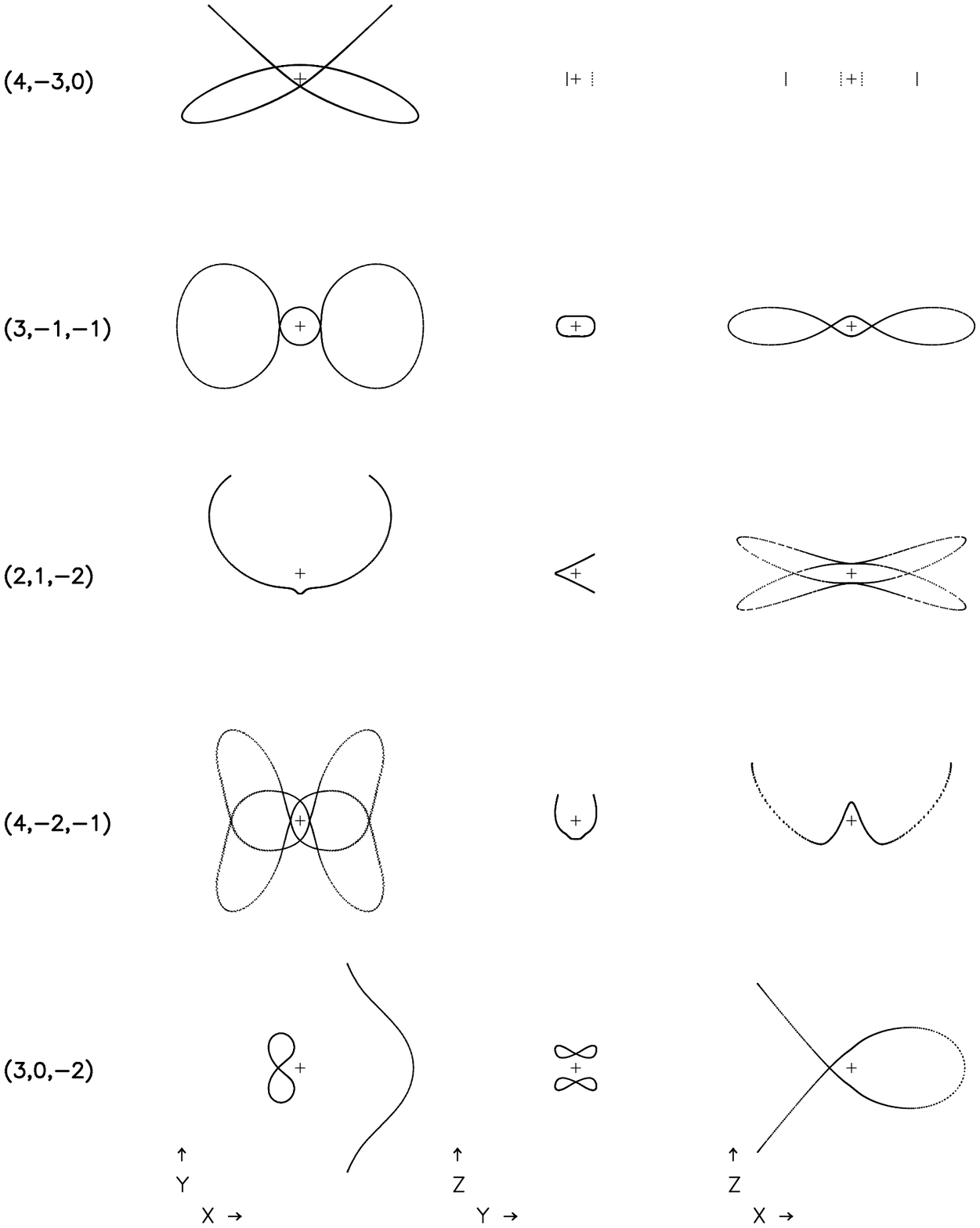]{\label{fig4}} 
Intersections of six thin box orbits from the potential of Figure 
3a with the principal planes.
Because the orbits are thin, their intersections with any plane 
define a curve or set of curves.
Each of these orbits is stable and avoids the center, whose 
position is indicated with a cross.

\figcaption[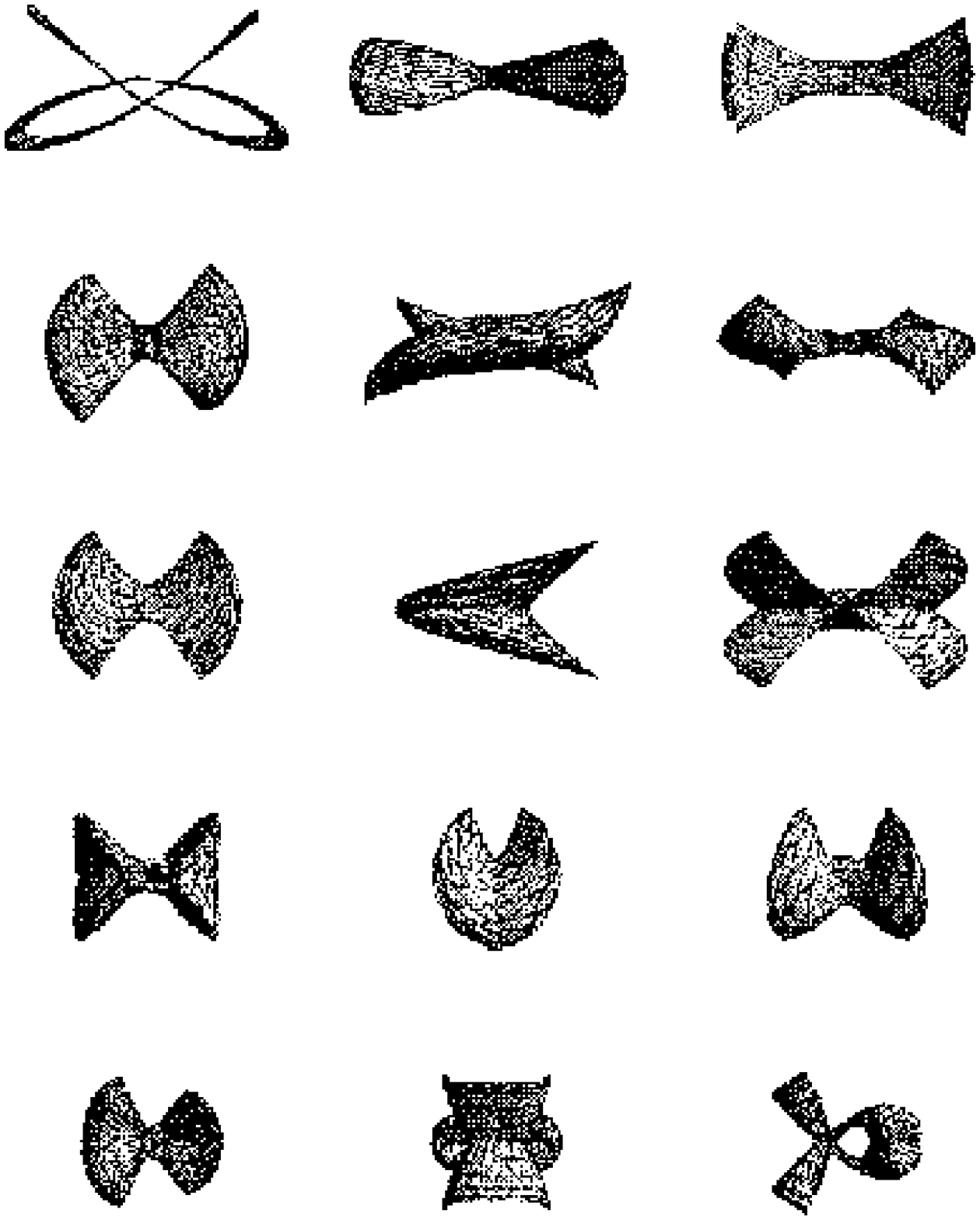]{\label{fig5}} 
Plots of the surfaces filled by the thin box orbits whose cross 
sections are shown in Figure 4, as seen from vantage points 
on each of the three principal axes.

\figcaption[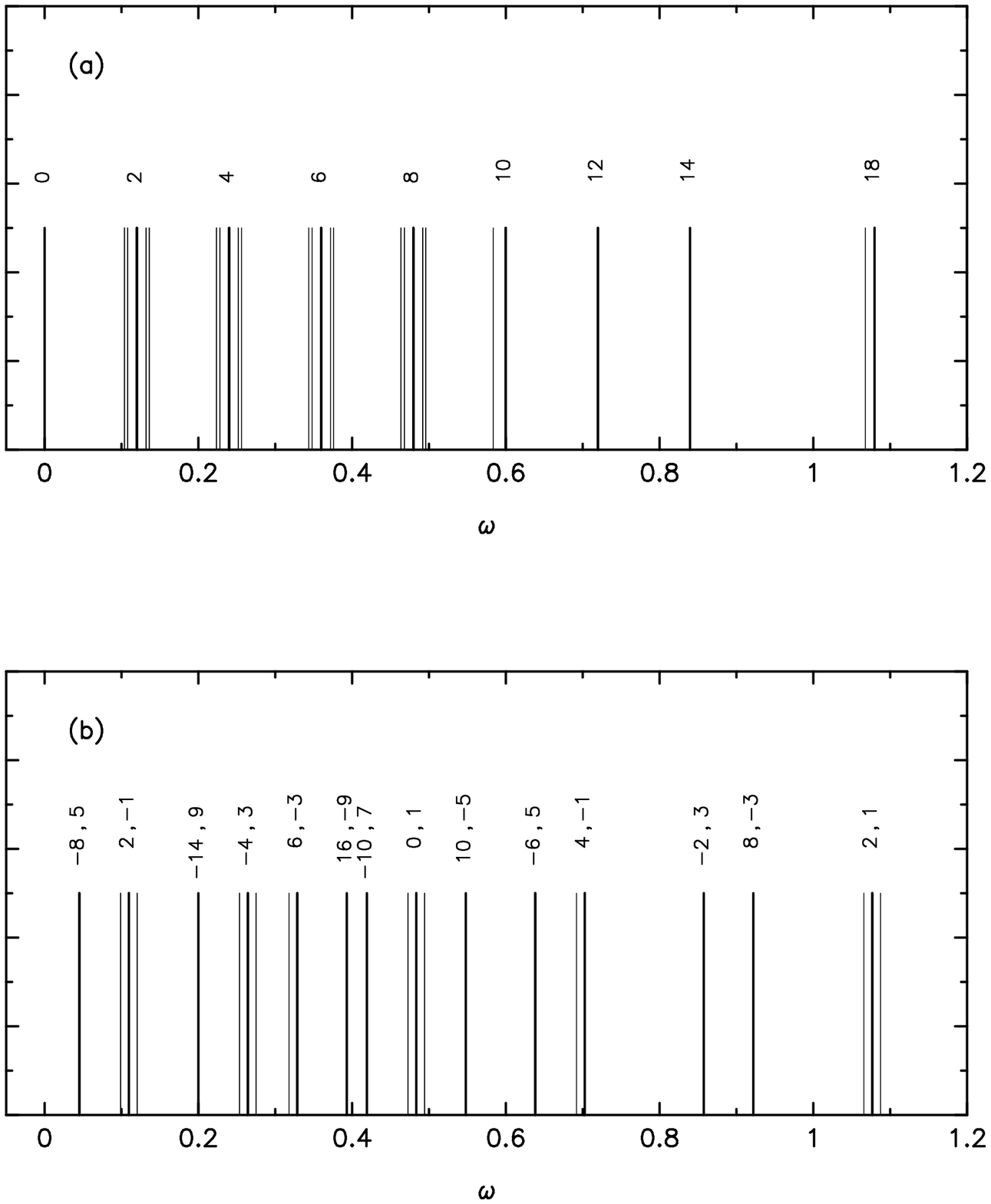]{\label{fig6}}
Schematic frequency spectra of the $z$-motion of two non-resonant
orbits from the potential of Figure 3a.
(a) An orbit that lies close to the $5:6:8$ closed orbit;
(b) an orbit that lies close to a thin orbit from the $(2,1,-2)$ 
resonance zone.
The amplitudes of the frequency spikes have been set to a 
constant value and only the absolute values of the frequencies
are plotted.
The ``primary'' lines of each spectrum are plotted in bold.
In (a), these lines lie at integer multiples of $\omega_0^{(1)}$, 
as indicated.
In (b), the primary lines lie at $\omega_k=l_k\omega_0^{(1)} 
+ m_k\omega_0^{(2)}$; each such line is labelled by $l_k$ and 
$m_k$.
The majority of regular orbits in the potentials investigated here
have spectra similar to (b), 
indicating that they are associated with a thin (singly-degenerate), 
rather than with a closed (doubly-degenerate), orbit.

\figcaption[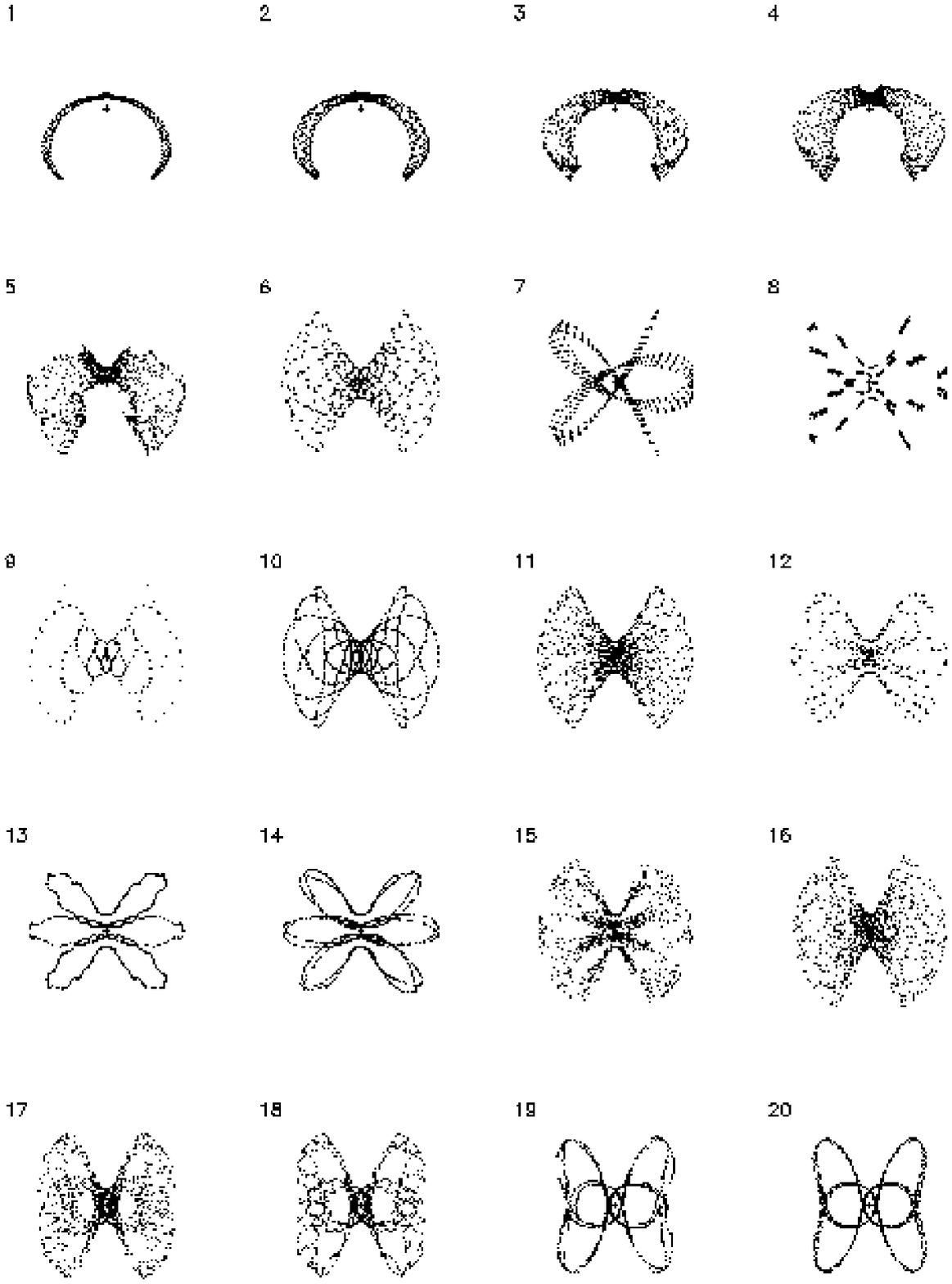]{\label{fig7}}
$x-y$ cross sections of a set of orbits whose starting points lie 
equally-spaced on the heavy line of Figure 3a, between the 
$(2,1,-2)$ and $(4,-2,-1)$ resonance zones.
This potential has no central ``black hole'' and the cusp is
weak, $\gamma=0.5$, making the motion nearly integrable.
Orbit 7 lies close to the $(-5,4,0)$ resonant orbit, 
and orbit 13 is close to the $(1,3,-3)$ resonance.
Many of the other orbits (e.g. nos. 6, 9, 10)
can be assigned to higher-order resonances.
Orbits 5, 11 and 16 are weakly stochastic.

\figcaption[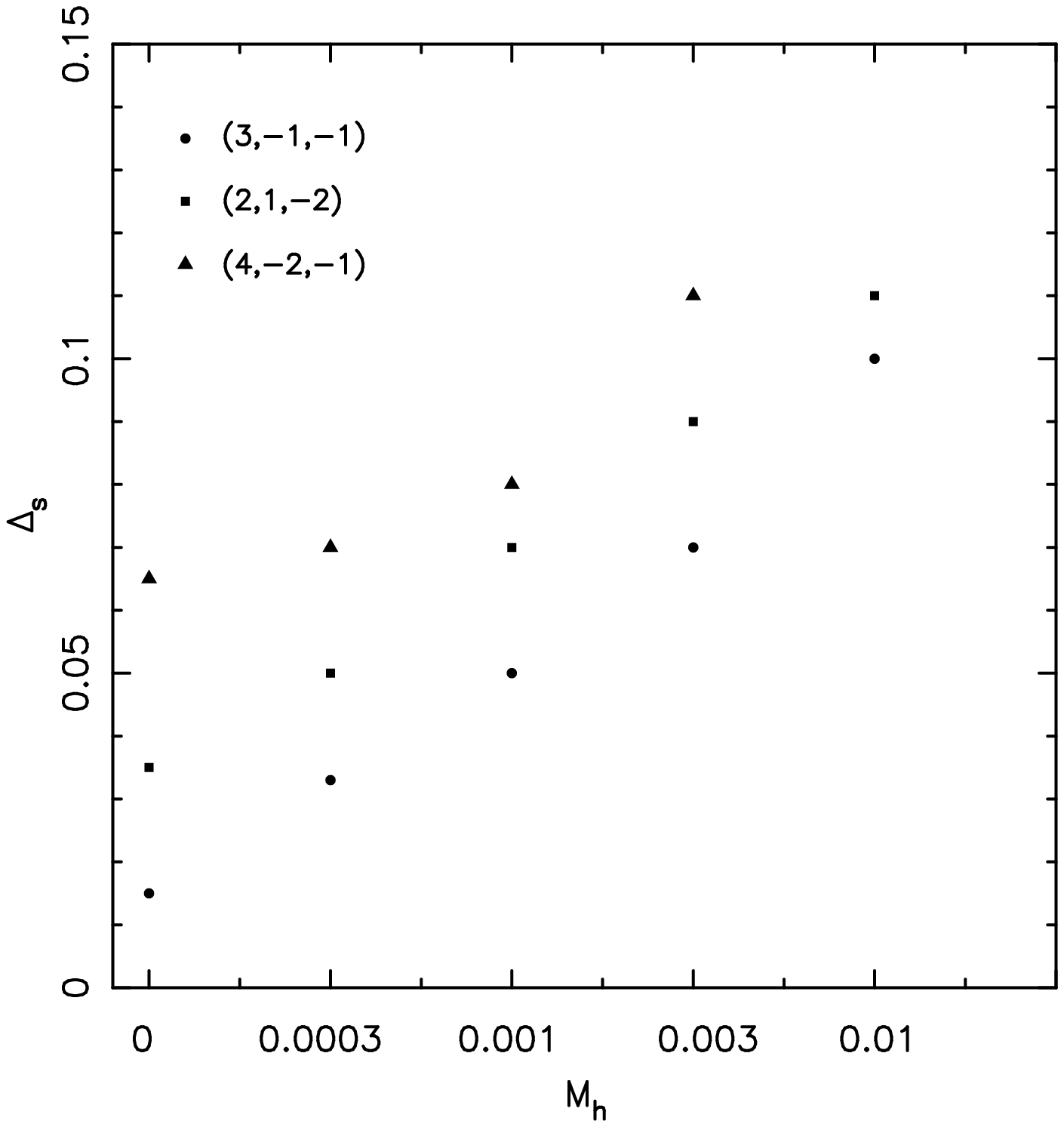]{\label{fig8}} 
Approximate pericenter distance $\Delta_s$ at which orbits from 
three resonant families become stochastic, as a function of the 
mass $M_h$ of a central black hole.
$\Delta_s$ was found by moving along a line of initial conditions 
like that of Figure 3b and recording the minimum pericenter 
distance at which orbits from the resonant family were regular.
The amplitude of the long-axis orbit is roughly 2 in each of these 
model potentials.
For $M_h\gap 0.01$, the $(4,-2,-1)$ family is fully stochastic;
the other two families are stochastic for $M_h\gap 0.03$.

\figcaption[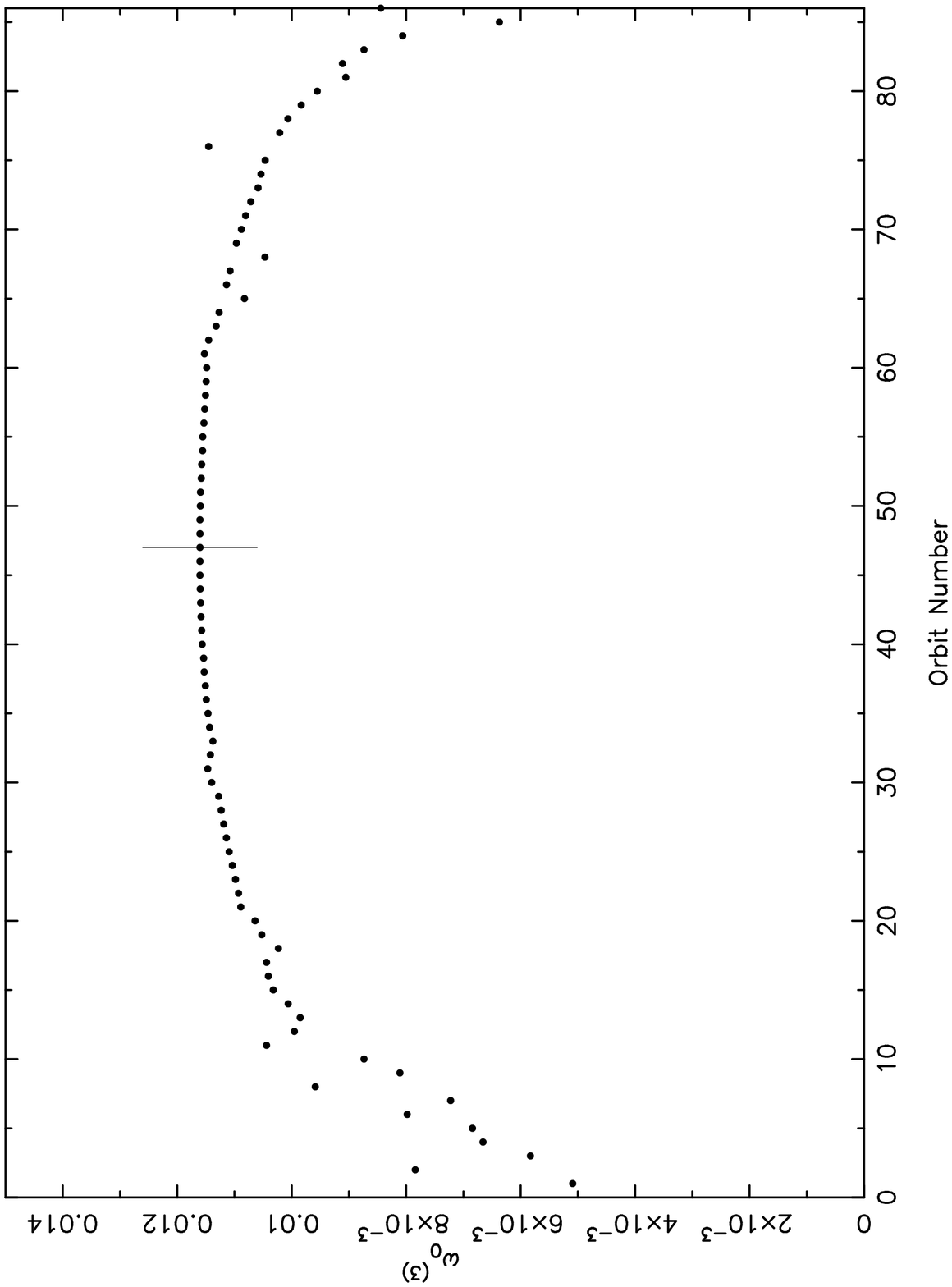]{\label{fig9}} 
Variation of the splitting frequency $\omega_0^{(3)}$ with 
starting point as one moves across the $(2,1,-2)$ resonance zone 
in the potential with $\gamma=0.5$ and $M_h=0.0003$.
The position of the resonant orbit is indicated by the thin 
vertical line.
Most orbits with numbers lying between 10 and 80 are regular.
Discontinuities occur when a second resonance condition is satisfied,
producing a closed orbit.

\figcaption[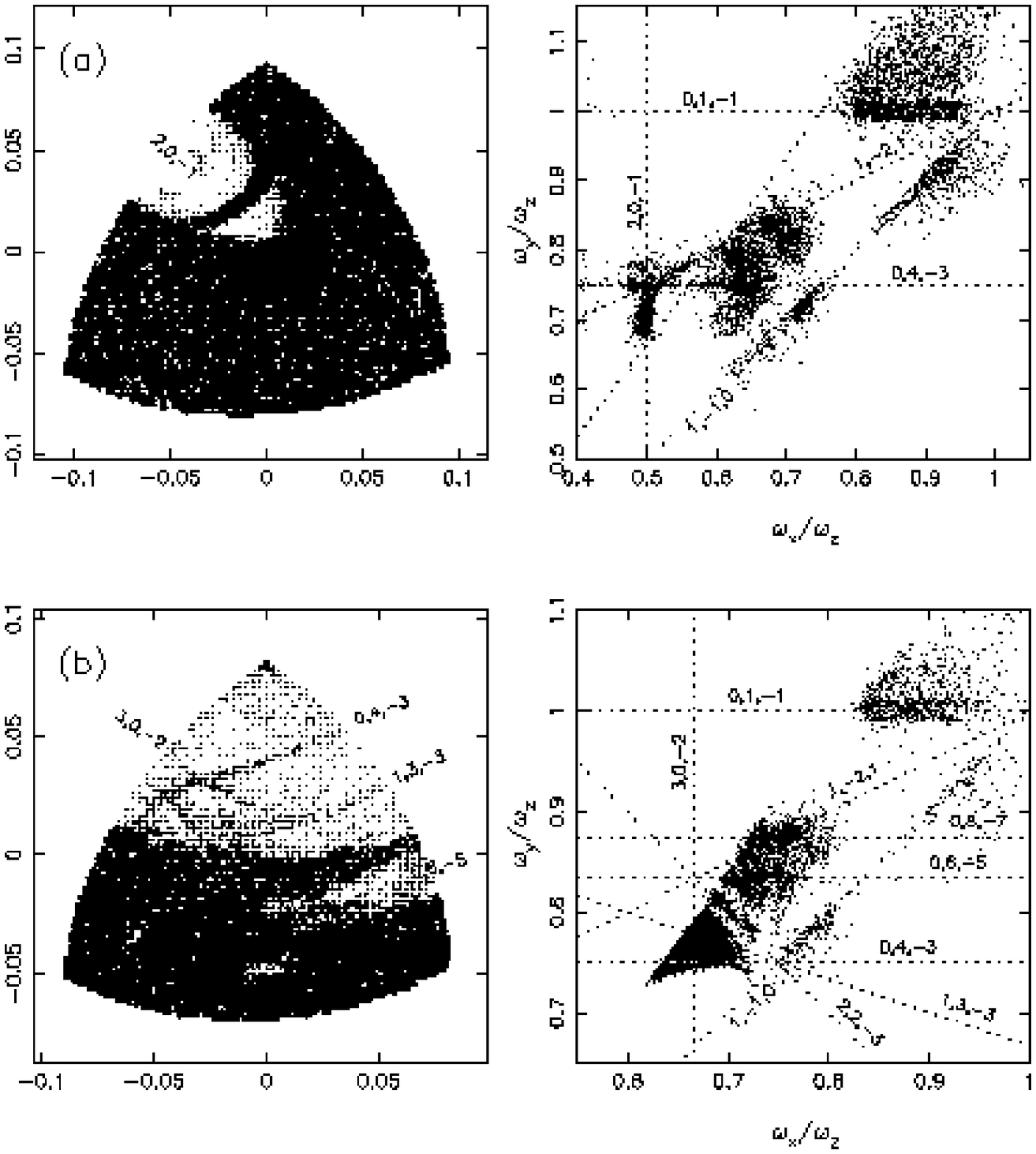]{\label{fig10}} 
Properties of boxlike orbits near the central black hole in
a triaxial galaxy.
The potential is that corresponding to a
Dehnen model with $\gamma=0.5$ and $M_h=0.003$.
Left panels show one octant of the equipotential surface
with the grey scale indicating the degree of stochasticity.
The top, left and right corners correspond to the $z$ (short),
$x$ (long) and $y$ (intermediate) axes.
Right panels are the frequency maps.
The most important resonance zones are labelled with their 
defining integers $(m_1,m_2,m_3)$.
(a) Shell $J=7$ ($M_h/M_J=0.189$); 
(b) shell 6 ($M_h/M_J=0.221$); 
(c) shell 5 ($M_h/M_J=0.265$); 
(d) shell 4 ($M_h/M_J=0.331$).

\figcaption[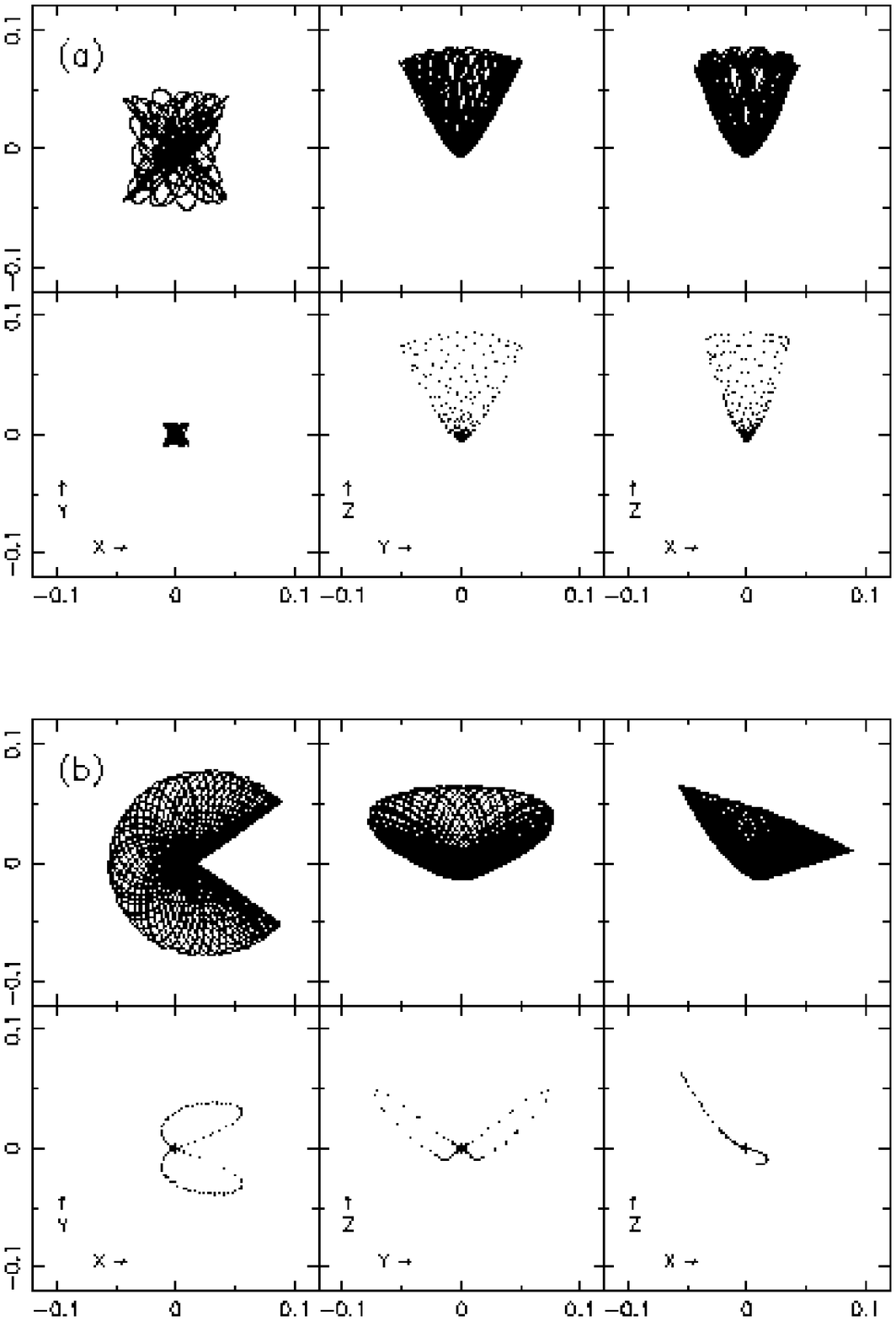]{\label{fig11}} 
Non-resonant (a) and resonant (b) orbits near the center of a
triaxial galaxy containing a nuclear black hole.
The orbit in (b) is associated with the $(1,-2,1)$ resonance.
Both orbits are taken from the set whose properties are displayed 
in Figure 10d.
Top panels are projections along the principal axes and
bottom panels show intersections with the three principal planes.
The non-resonant orbit (a) is volume-filling while the resonant
orbit (b) is thin.
The position of the black hole is marked with a cross.

\clearpage

\setcounter{figure}{0}

\begin{figure}
\plotone{fig1.ps}
\caption{ }
\end{figure}

\begin{figure}
\plotone{fig2.ps}
\caption{ }
\end{figure}

\begin{figure}
\plotone{fig3.ps}
\caption{ }
\end{figure}

\begin{figure}
\plotone{fig4.ps}
\caption{ }
\end{figure}

\begin{figure}
\plotone{fig5.ps}
\caption{ }
\end{figure}

\begin{figure}
\plotone{fig6.ps}
\caption{ }
\end{figure}

\begin{figure}
\plotone{fig7.ps}
\caption{ }
\end{figure}

\begin{figure}
\plotone{fig8.ps}
\caption{ }
\end{figure}

\begin{figure}
\plotone{fig9.ps}
\caption{ }
\end{figure}

\begin{figure}
\plotone{fig10a.ps}
\caption{ }
\end{figure}

\setcounter{figure}{9}

\begin{figure}
\plotone{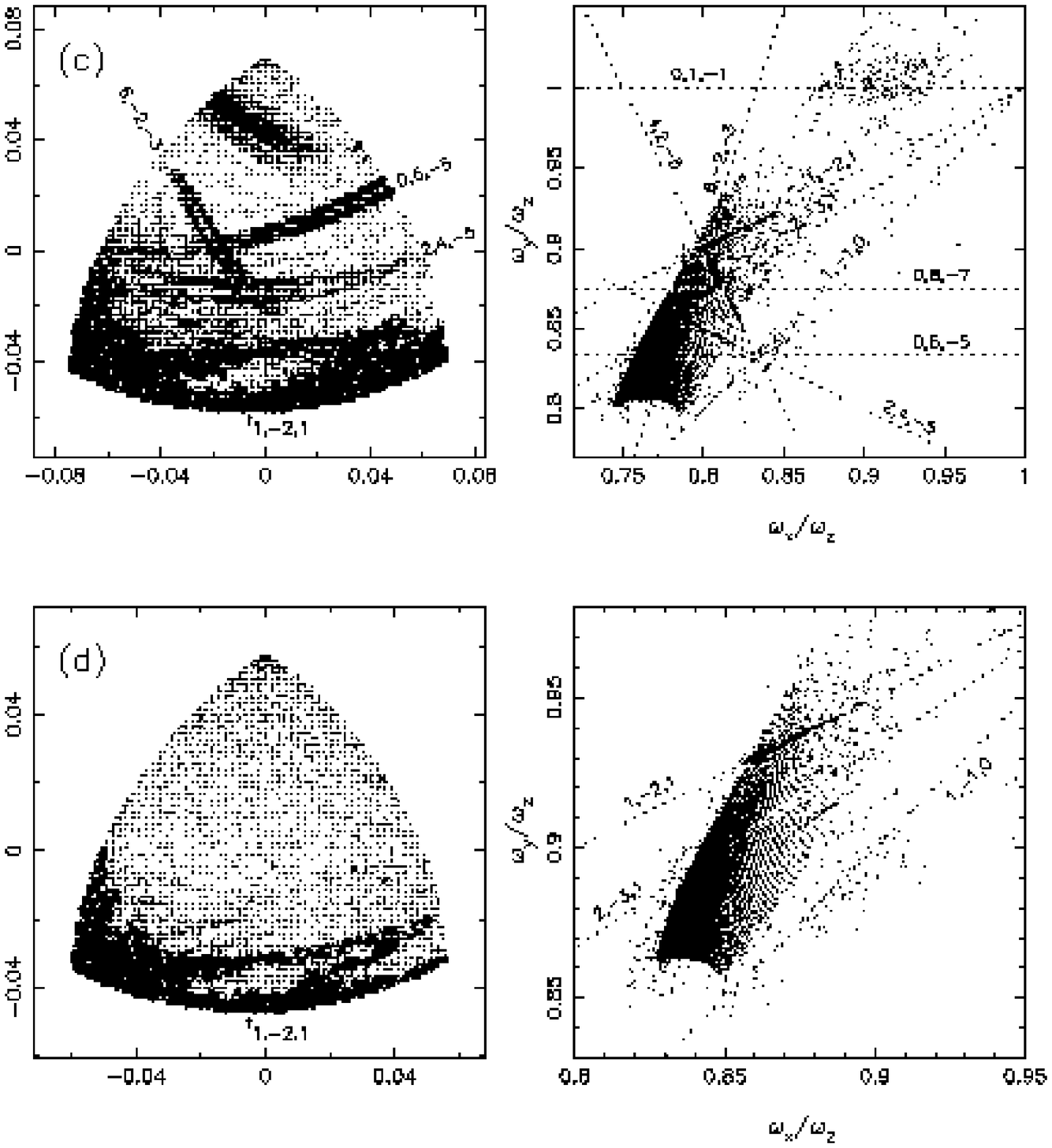}
\caption{ }
\end{figure}

\begin{figure}
\plotone{fig11.ps}
\caption{ }
\end{figure}

\end{document}